\documentclass[prl,twocolumn,english,superscriptaddress]{revtex4-1}

\usepackage[T1]{fontenc}
\usepackage[latin9]{inputenc}
\usepackage{color}
\usepackage{babel}
\usepackage{latexsym}
\usepackage{float}
\usepackage{amsmath}
\usepackage{amsfonts}
\usepackage{graphicx}
\usepackage{times}   
\usepackage{esint}
\usepackage{subfigure}
\usepackage{verbatim}
\usepackage{algorithm}
\usepackage{algorithmicx}
\usepackage{algpseudocode}
\usepackage{soul}
\usepackage{xcolor,colortbl}
\usepackage[unicode=true,pdfusetitle,
 bookmarks=false,colorlinks=true,citecolor=blue,urlcolor=blue,linkcolor=red]{hyperref}
\newcommand{\bra}[1]{\left\langle #1 \right|}
\newcommand{\ket}[1]{\left|#1\right\rangle}
\newcommand{\braket}[2]{\left\langle#1 |  #2\right\rangle}

\makeatletter

\newcommand{\mc}[2]{\multicolumn{#1}{|c|}{#2}}
\@ifundefined{textcolor}{}
{%
 \definecolor{BLACK}{gray}{0}
 \definecolor{WHITE}{gray}{1}
  \definecolor{WHITE1}{gray}{0.99}
 \definecolor{RED}{rgb}{1,0,0}
 \definecolor{GREEN}{rgb}{0,1,0}
 \definecolor{BLUE}{rgb}{0,0,1}
 \definecolor{CYAN}{cmyk}{1,0,0,0}
 \definecolor{MAGENTA}{cmyk}{0,1,0,0}
 \definecolor{YELLOW}{cmyk}{0,0,1,0}
 \definecolor{LightCyan}{rgb}{0.78,1,1}
}

\@ifundefined{date}{}{\date{}}
\AtBeginDocument{
  
}
\makeatother

\newcommand{\SAVE}[1]{}

\begin{document}
\renewcommand\abstractname{}

\title{An efficient deterministic perturbation theory for selected configuration interaction methods}
\author{Norm M. Tubman, Daniel S. Levine, Diptarka Hait, Martin Head-Gordon, K. Birgitta Whaley}
\affiliation{Kenneth S. Pitzer Center for Theoretical Chemistry, Department of Chemistry, University of California, Berkeley, California 94720, USA and Chemical Sciences Division, Lawrence Berkeley National Laboratory Berkeley, California 94720, USA}
\date{\today}
\begin{abstract} 
The interplay between advances in stochastic and deterministic algorithms has recently led to development of interesting new selected configuration interaction (SCI) methods for solving the many-body Schr\"{o}dinger equation. The performance of these SCI methods can be greatly improved with a second order perturbation theory (PT2) correction, for which  stochastic and hybrid-stochastic methods have recently been proposed as new tools to perform such calculations. 
In this work, we present a highly efficient, fully deterministic PT2 algorithm for SCI methods and demonstrate that our approach is orders of magnitude faster than recent proposals for stochastic SCI+PT2. We also show that it is important to have a compact reference SCI wave function, in order to obtain optimal SCI+PT2 energies. This indicates that it  advantageous to use accurate search algorithms such as 'ASCI search' rather than more approximate approaches. Our deterministic PT2 algorithm  is based on sorting techniques that have been developed for modern computing architectures and is inherently straightforward to use on parallel computing architectures.  Related architectures such as GPU implementations can be  also used to  further increase the efficiency. Overall, we demonstrate that the algorithms presented in this work allow for efficient evaluation of trillions of PT2 contributions with modest computing resources. 
\end{abstract}
\maketitle
\newpage


\section{Introduction}
\label{sec:intro}

There has been a recent renaissance in the field of selected configuration interaction (SCI) algorithms \cite{tubman2016-1,evangelista2014,knowles2015,holmes2016,evan2016,wenjian2016,evan2017,garniron2017,tubman2018} for tackling strongly correlated systems where traditional single reference theories like coupled cluster methods are likely to break down. 
Recently, SCI has been applied to systems that are larger than what is possible with conventional FCI algorithms~\cite{siegbahn1984,knowles1984,olsen1988,sherrill1999,gan2005}.  This include simulations of transition metal systems~\cite{tubman2016-1,holmes2016}, a solver for dynamical mean field theory~\cite{zaera2017}, cluster decompositions~\cite{lehotola2017}, and excited states~\cite{tubman2016-1,evan2017,loos2018mountaineering}.
One approach to SCI is called adaptive sampling configuration interaction (ASCI)~\cite{tubman2016-1,tubman2018}, which has been developed with fast and efficient algorithms to take advantage of modern computing architectures.   

SCI methods typically include a second order perturbation theory (PT2) correction to account for the relatively small component of the correlation energy missed by the variational SCI wave function\cite{bender1969,evangelisti1983,harrison1991}. Indeed, it was also observed that the accuracy of ASCI can be increased dramatically by incorporation of a PT2 correction \cite{tubman2016-1}. 
In very recent years, stochastic perturbative corrections have been developed for various SCI methods~\cite{garniron2017,sharma2017}, as well as for density matrix renormalization group (DMRG) \cite{guo2018-2} and for FCIQMC~\cite{blunt2018communication}.
The apparent lack of fast deterministic PT2 algorithms in the literature  has perhaps encouraged development of these recent stochastic approaches.  Two of the newer stochastic PT2 developments for SCI are the 
semi-stochastic perturbation theory~\cite{sharma2017} and hybrid deterministic-stochastic~\cite{garniron2017} algorithms.  

It is an interesting and still open question as to whether or not stochastic PT2 algorithms have inherent advantages over their deterministic counterparts. 
To address this question, in this work we introduce a fast deterministic approach to generate exact PT2 energies. We also demonstrate via comparison of run times, that this new deterministic algorithm is significantly faster than the recently proposed stochastic algorithms for SCI.

\section{Overview of ASCI}
\label{sec:overview}

\begin{figure}
\begin{center}
\scalebox{1}{\includegraphics[width=1.0\columnwidth]{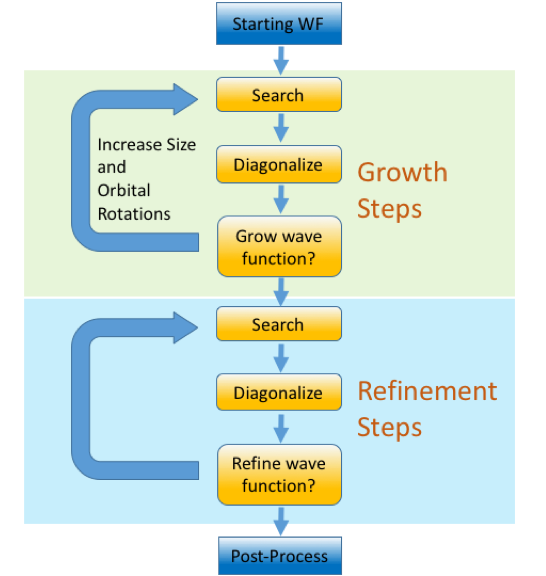}}
\end{center}
\caption{A flowchart of the ASCI algorithm.  The main computational components are the search, diagonalization, and post-processing steps.  
The growth steps are done in the first set of iterations of ASCI to bring the variational wave function from the Hartree-Fock determinant to a wave function of size $N_{tdets}$.  We grow the wave function in steps, since we find this to be faster than repeated search/diagonalization iterations starting from an inaccurate full-sized initial wave function.   The size of the wave function is held fixed during the refinement steps, but the quality is improved via search/diagonalization iterations. This helps generate a highly accurate variational wave function. The post processing step generally improves the energy further beyond the variational ASCI wave function.  This step can be carried out with perturbation theory as described in this work, or with methods such as auxiliary field quantum Monte Carlo~\cite{Motta_WIREs_2018} or diffusion Monte Carlo~\cite{tubman2011,qmcpack2018}.}
\label{fig:flowchart}
\end{figure}
 
The main idea behind SCI methods (see Figure~\ref{fig:flowchart} for a flowchart of the ASCI approach) is to diagonalize the Hamiltonian within a Hilbert space in which only the most important degrees of freedom are identified and retained.  SCI algorithms therefore attempt to generate the top contributing determinants to the full CI (FCI) wave function, in contrast to traditional FCI algorithms, which include all possible determinants. They also stand in contrast to truncated CI approaches, like CISD, CISDT, etc, which typically consider restricted excitations from the reference up to some given level of excitation\cite{pople1977,sherrill1999}.
Truncation based upon excitation level leads to large size-consistency errors~\cite{pople1977}, that make such models unsuitable for most chemical applications, while SCI approaches like ASCI simply aim for a sufficiently close approximation to FCI that this issue is irrelevant.

Full details of the ASCI method are presented in previous works~\cite{tubman2016-1,tubman2018}, and we therefore provide only a brief overview here.  In the ASCI method, a wave function ($\psi_{k}$) is improved upon over the course of several iterations.
In each iteration $k$, the current wave function and associated Hamiltonian matrix elements are used together to search the Hilbert space for new determinants.
The search part of the algorithm requires two rules: a selection criterion to determine what part of Hilbert space to search (pruning), and a ranking criterion over the determinants that make it through the pruning, to determine the best determinants to include in the updated wave function $\psi_{k+1}$.  Two different pruning techniques have been discussed in the original ASCI and heatbath CI (HCI) papers: these we refer to as coefficient driven search~\cite{tubman2016-1} and integral driven search~\cite{holmes2016}, respectively.  The original HCI ranking algorithm however, is quite different and more approximate than the one used in ASCI (as will be shown later, this reduces the relative compactness of the variational wave functions).  The full algorithm in ASCI is simply called ASCI search and is described in detail in recent work~\cite{tubman2018}.  

The ranking criterion that is most widely used is derived from a consistency relationship among the coefficients of the determinants that constitute eigenstates of the Schr\"{o}dinger equation.   Specifically, the consistency equation states that if  we consider an expansion of an eigenstate in a basis of Slater determinants with coefficients $C_i$, we will have:
\begin{equation}
C_{i} = \frac{\sum_{j \ne i}H_{ij}C_{j}}{(E-H_{ii})},
\end{equation} 
\noindent where $H_{ij}$ is the Hamiltonian matrix element between the $i$th and $j$th determinant, and $E$ is the energy of the eigenstate.
This equation can be reinterpreted as an iterative recipe for obtaining a better set of determinants for expanding $\psi_{k+1}$ by feeding in as input an approximate wave function $\psi_k$.  This is done by finding improved 
coefficients $A_i$ from the input coefficients $C^k_{j}$, according to
\begin{equation}
A_{i} = \frac{\sum_{j \ne i}H_{ij}C^k_{j}}{E_k-H_{ii}}.
\label{eqn:rankeqn}
\end{equation}
\noindent where $C^k_j$ is the CI expansion coefficient of the $j$th determinant in the $k$th wave function iteration and $E_k$ is the energy of the $k$th wave function $\psi_k$. Thus, $C^{k}_i$ are the coefficients of the input wave function, and the output coefficients $A_{i}$ are estimates of coefficients of an improved wave function that is closer to an eigenstate. The $A_i$ coefficients are also related to a first-order perturbation estimate for CI coefficients in many body Epstein-Nesbet perturbation theory ~\cite{huron1973}.  

Since the goal of a SCI method is to include the most important weight determinants in the expansion, ASCI search uses a slightly modified version of Eqn.~\ref{eqn:rankeqn} to define a ranking, where $|A_i|$ is the rank value of the $i$th determinant.  The novel algorithm used in ASCI search to calculate the ranking equation is both fast and accurate.  It is described in detail in reference~\cite{tubman2018}.  
After the ranking values are calculated, the top $N$ determinants are then chosen for Hamiltonian construction and subsequent diagonalization to obtain $\psi_{k+1}$ and $E_{k+1}$.  In practice, this iterative approach  is successful in generating all the top contributions to the wave function, which is essential for obtaining highly accurate energies. The effect of the neglected determinants can be approximated via perturbation theory~\cite{tubman2016-1}.  Additional considerations for performing other aspects of selected CI, such as Hamiltonian building, have also been described in recent work~\cite{tubman2018}.

The PT2 energy correction $E_{PT2}$ is generally calculated with Epstein-Nesbet perturbation theory~\cite{epstein1926,nesbet1955}, in which it is given by 
 \begin{equation}
 E_{PT2} = \sum_{i}\frac{|\langle\psi|H|D_{i}\rangle|^{2}}{E_{ASCI}-H_{ii}}. \label{eqn:en}
 \end{equation}
This combination of ASCI search with Epstein-Nesbet PT2 is able to treat strongly correlated systems, including the Cr$_{2}$ dimer in the SVP basis, to chemical accuracy~\cite{tubman2016-1}. To the best of our knowledge, ASCI (as well as other SCI methods) and DMRG approaches \cite{yanai2009,amaya2015} are the only methods that have been able to treat Cr$_{2}$ to chemical accuracy in the SVP basis.  We note that 
it is possible that auxiliary field quantum Monte Carlo will soon also reach this benchmark, using the new and improved wave functions that are currently under investigation~\cite{borda2018}.

\section{Deterministic Perturbation Theory with ASCI}
\subsubsection{The Fundamental Algorithm}
Understanding modern computer algorithms and architectures is important in designing new simulation tools, since optimal performance depends on their interplay.  
To start developing an efficient formulation of the PT2 step on modern computers, we first introduce what we consider to be the fundamental algorithm for Epstein-Nesbet PT2 in this section (also summarized in Algorithm~\ref{alg:pt2}). 

The algorithm begins with generation of all possible PT2 connections. These are stored in an array (as bitstrings) together with their PT2 contribution.  The main part of the algorithm is to sort this array by the bitstring value.  The sorted array is now arranged such that the Epstein-Nesbet PT2 energy (Eq.~\ref{eqn:en}) can be calculated with a single pass through the contributions. 
This approach is conceptually simple and easy to implement. We suggest that it is likely among the fastest algorithm that can be performed on modern computers, with the caveat that large amounts of memory would be needed to store all of the contributions.  

The efficiency of this approach stems from the fact that sorting algorithms have been extensively optimized on modern computers, due to their ubiquitous appearance in all aspects of computation~\cite{gpusort2017}.  The minimization of cache misses on modern computers are likely among the main reasons that sorting approaches 
appear to be significantly more efficient than other algorithms~\cite{tubman2018,ips4,pdq}, although many aspects of sorting algorithms have been designed for optimal usage on modern computers.  Hash tables are an example of another structure that can be used to accomplish the same task, but they do not appear to be as efficient in our tests using standard libraries (such as STL and Boost) ~\cite{boost2011,plauger2000}. Recent benchmarks with state of the art hash tables suggest that there still may be a possibility of this becoming as efficient as the sorting approach for applications of selected CI~\cite{hash_benchmark}.   Many different sorting algorithms have been benchmarked with bitstrings generated within an ASCI simulation in recent work~\cite{tubman2018}.

\begin{figure}[tpb]
\begin{algorithm}[H]
 \begin{algorithmic}[1]
\State Start with a wave function $\psi$ for which to calculate the PT2 energy
\State Generate all PT2 contributions and store their bitstring and PT2 contributions as follows:
\begin{itemize}
\item  Generate PT2 contributions, $|D_{j}\rangle$, by acting the Hamiltonian on each determinant in $\psi$, $H|D_{i}\rangle$
\item  Store the bitstring associated with $|D_{j}\rangle$, along with the numerator of the PT2 contribution $H_{ij}C_{i}$
\end{itemize}
\State Sort all PT2 contributions by bitstring
\State Calculate equation~\ref{eqn:en} by looping over the sorted PT2 contributions as follows:
\begin{itemize}
\item For each unique bitstring $j$ not currently in $\psi$:  $E_{j}$ = $\sum_{i}{\frac{H_{ij}C_{i}}{E_{ASCI}-H_{jj}}}$ 
\item Update the PT2 energy: $E_{PT2}$ = $E_{PT2}$+ $E_{j}$
\end{itemize}
 \end{algorithmic} 
 \caption{Fundamental Algorithm for Epstein-Nesbet PT2}
 \label{alg:pt2}
\end{algorithm}
\end{figure}

\subsubsection{Constraint-based PT2 }
The fundamental algorithm for calculating Epstein-Nesbet PT2 (Algorithm~\ref{alg:pt2}) is efficient for small systems in which the entire set of PT2 corrections fit into memory. This is however not the case for larger systems, since the number of possible PT2 correction terms grows rapidly with system (and basis) size.  
Here we present a constraint-based PT2 approach for large systems.  The constraint-based approach has a key algorithmic improvement over Algorithm~\ref{alg:pt2} that allows it to overcome memory bottlenecks and to become embarrassingly parallel, with near linear scaling over the number of processors.  We present several results obtained with this approach and demonstrate that enormous speedups are possible in comparison to stochastic approaches~\cite{garniron2017,sharma2017}.  We then describe in detail the related algorithmic advances that we developed in order to make the algorithm as fast as possible. 

The constraint-based PT2 approach works by generating PT2 contributions in batches that satisfy a constraint which partitions the entire space of contributions into \textit{non-overlapping} subsets. 
These non-overlapping subsets allow the PT2 calculation to proceed in a trivially parallelizable manner, with each thread/processor assigned its own subset.
This situation is also called embarrassingly parallel in the literature.
 The parallelization of constraints is easy to set up, with only the complication of implementing load balancing to prevent asymmetric distribution of tasks across threads/processors. 
The generic constraint PT2 algorithm is summarized in Algorithm~\ref{alg:constraint}.

\begin{figure}[tpb]
\begin{algorithm}[H]
 \begin{algorithmic}[1]
\State Start with a wave function $\psi$ for which to calculate the PT2 energy
\State Generate a set of non-overlapping and complete set of constraints
\State Loop over constraints
\begin{itemize}
\State Generate all PT2 contributions subject to the current constraint and store their bitstring and PT2 contributions
\State Sort all current PT2 contributions by bitstring
\State Calculate equation~\ref{eqn:en} by looping over the sorted PT2 contributions
\item Add the energy from the current constraint to the total PT2 correction
\end{itemize}
 \end{algorithmic}
 \caption{Constraint Epstein-Nesbet PT2}
 \label{alg:constraint}
\end{algorithm}
\end{figure}

\subsubsection{Triplet constraints}
The constraint approach is quite general in terms of the possible constraints that can be used.   The implementation we present here uses what we call the \textit{triplet constraints}.  These are indexed by a set of three numbers that specify the three highest occupied alpha spin-orbitals in the Slater determinant for a specific PT2 contribution.  These three numbers uniquely specify a class of determinants (PT2 contributions), and thus the triplet constraints are non-overlapping and complete.  For our purposes, we order the bitstrings such that the alpha bitstrings are the most significant digits (orbitals).  Thus in any system in which there are at least three or more alpha electrons, the triplet values will refer to the occupation of the top three alpha electrons.  

This approach is summarized in Algorithm \ref{alg:constraint}. It works by first constructing a loop over the triplet constraints.  For each triplet constraint, we then consider each determinant in the variational ASCI wave function and generate all possible PT2 contributions (i.e., single and double excitations) from this that satisfy the given triplet constraint. 
These contributions are stored in an array and then sorted after all contributions have been generated.  This aggregates all the terms that make up all PT2 contributions with the given constraint. A single pass over the sorted list is then sufficient for calculating the PT2 energy contribution with Eqn.~\ref{eqn:en}. 

The motivation for the triplet constraints comes from specific consideration of the simulations we are targeting.  For systems with around 50 electrons, 200 orbitals, and up to 10 million determinants in the SCI reference wave function,  we find that in most cases the largest number of contributions per constraint is 1 million to 100 million, which can easily fit into memory on most modern machines.  Larger systems are also likely to fit in memory. In the instances where this is not the case, it is easy to modify the constraints, e.g., by specifying the top four orbitals (quadruplet constraint) instead of the top three.  This flexibility leads to easy generation of memory and speed efficient approaches.

\subsubsection{ Parallelization}
The algorithm that we have outlined above is straightforward to parallelize. 
While the triplet constraint described above shows a large variation in the number of PT2 contributions for each constraint, the load balancing issue can nevertheless be resolved by creating a work-list with an accurate estimate of the number of contributions in each constraint and distributing the work accordingly.   Due to the flexibility of our constraints, specifically, the fact that each constraint can be broken up into subconstraints, load balancing becomes a matter of implementation and presents no inherent algorithmic difficulty.   Furthermore, for a given constraint, the computational work can be parallelized, as most of the computational time is focused on generation of the PT2 contributions and sorting of these contributions.  Thus openmp, MPI, and openmp/MPI hybrid approaches can be used straightforwardly.   To be explicit, our ASCI implementation on 20 cores is perfectly parallel, and can be directly compared to the 20 core SHCI simulations included in the results section.   We present single core timings in the results section to make it easier to compare to other works for which benchmarks are presented for different numbers of processors.

\section{Results}
In this section, we present results for implementation of the triplet constraint evaluation of PT2 corrections to ASCI.  We use a cutoff 
which ignores PT2 contributions (given by the numerator part $|c_{i}H_{ij}|$, where $i$ is in the variational space and $j$ is the perturbing determinant) that are less than 10$^{-8}$ ~\cite{holmes2016}. This is not strictly necessary, since the deterministic approach can handle all the contributions efficiently with only a reasonable amount of extra computational effort, at least for the systems tested in this work.  To know whether to include a term we currently calculate $|c_{i}H_{ij}|$ for all contributions, however for terms that are neglected we save time by not calculating the denominators or sorting such terms.  When one is interested in developing a lower accuracy deterministic PT2 approach, integral driven versions of this algorithm can be created and used to more efficiently neglect terms without having to calculate the numerator part explicitly~\cite{holmes2016}. We find the errors on the total energy that result from these neglected contributions are generally less than 10$^{-7}$ Hartrees for the systems considered in this work. 
Additionally, neglecting these terms allows us to make explicit comparisons to previously published HCI results.

In Tables~\ref{tab:dz},~\ref{tab:tz}, and~\ref{tab:qz} we demonstrate the performance of our approach using canonical Hartree-Fock orbitals,  making comparison to the stochastic PT2 method used with HCI~\cite{sharma2017} (which refered to as SHCI).  In Table~\ref{tab:qz} we also make comparison to the deterministic-stochastic hybrid PT2 implemented with CIPSI~\cite{garniron2017}.  To compare deterministic results against stochastic results, we calculate the amount of time it takes to achieve an error bar of 0.1 mHa accuracy with a 95\% likelihood.  This would typically be called a $2\sigma$ error bar in the Monte Carlo literature.  
 We also consider what error bars would be needed to achieve the same accuracy in an energy difference (since subtracting two quantities with the same error results in an error $\sqrt{2}$ larger than the original error).
Since error bars asymptotically decrease with the amount of sampling as  $\frac{1}{\sqrt{N_{samples}}}$, energy differences having the same error bars as the absolute energies require another factor of two greater computation time.  Since these algorithms are all inherently parallelizable, we calculate the single core times to compare against previously published results that use slightly different architectures.  We expect these comparison to be informative as all calculations are done on similar Intel processors, the details of which are described in the caption of the tables.   
\subsection{Molecules with Hartree-Fock orbitals}

We have analyzed C$_{2}$, N$_{2}$ and F$_{2}$, at internuclear separations of 1.24253 \AA, 1.0977 \AA, and 1.4119 \AA~respectively, for testing purposes. Table ~\ref{tab:dz} present timings for the triplet constraint approach compared with timings for the stochastic approaches.  Our results indicate that the deterministic constrained PT2 approach is in general two orders of magnitude faster than the stochastic PT2 approach employed in SHCI with the cc-pVDZ basis.  For cc-pVTZ and cc-pVQZ bases, the improvement over stochastic SHCI PT2 is between one and two orders of magnitude, as can be seen from Tables~\ref{tab:tz} and \ref{tab:qz}. In Table~\ref{tab:qz} we also make comparison to deterministic-stochastic hybrid CIPSI in the deterministic limit (i.e., in the limit that the stochastic part of the algorithm is turned off)~\cite{garniron2017}.

Our results show  that ASCI generates more \textit{compact} variational wave functions than HCI, which is expected (see also~\cite{tubman2018}).
By \textit{compact} we mean to make a comparison of the variational energy that can be attained for a given number of determinants.  Although ASCI and HCI search the same variational space, ASCI finds a significantly better set of determinants.   The only system in Table~\ref{tab:qz} that shows an exception to this is the case of F$_{2}$ in a cc-pVQZ basis, without natural orbital rotations. While the reason for the lower HCI energy in this case is not entirely clear, 
it is however evident that CIPSI is also not able to obtain the HCI variational energy, even though it also uses a search much improved over HCI.  
The variational energies for ASCI and CIPSI are comparable when the ASCI wave function has $2\times 10^{6}$ determinants and the CIPSI wave function has $4\times 10^{6}$. 

As can be seen in Tables~\ref{tab:dz},~\ref{tab:tz}, and~\ref{tab:qz} the ASCI wave functions are sometimes close to 40\% more compact than HCI  wave functions with the same accuracy of variational energy.  HCI appears unable to generate compact wave functions (as can be seen from variational energies in Tables \ref{tab:dz}, \ref{tab:tz} and \ref{tab:qz}), reflecting more significant residual systematic errors.  With the introduction of the fast ASCI search algorithm~\cite{tubman2018}, it is now feasible to do larger, more accurate searches to generate better reference wave functions, which not only produces better variational energies but also improved PT2 predictions.

\subsection{Molecules with Natural Orbital Rotations}
\begin{figure}
\begin{center}
\scalebox{1}{\includegraphics[width=1.0\columnwidth]{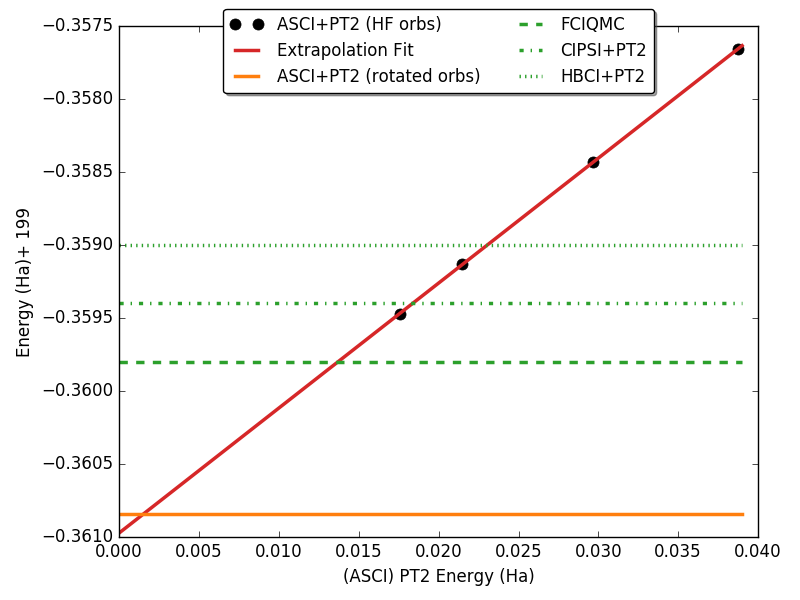}}
\end{center}
\caption{A comparison of F$_{2}$/cc-pVQZ  energies predicted by different methods.   The ASCI+PT2 results with Hartree-Fock orbitals appear to follow a linear trend vs the PT2 energy. Similar linear trends were also observed earlier for HCI~\cite{holmes2016} and CIPSI~\cite{loos2018mountaineering}. The extrapolated linear fit (using simulations with 10$^{5}$, 3*10$^{5}$, 10$^{6}$ and 2*10$^{6}$  determinants) gives an energy of -199.36097 Ha, which agrees very well with the -199.36082 Ha energy predicted by the best ASCI+PT2 result with natural orbital rotations (see natural orbitals results in Table~\ref{tab:orbrot}).   The best results from stochastic SHCI=-199.3590(9), CIPSI+PT2=-199.3594, and  FCIQMC=-199.3598(2) are plotted as with dotted and dashed lines, but all of these results are likely less converged than the ASCI results (see text for details).  The best results from stochastic SHCI have the same energy and error bars for simulations with and without orbital rotations.
}
\label{fig:f2}
\end{figure}

We now demonstrate the effect of adding orbital rotations for the F$_2$ molecule in Table~\ref{tab:orbrot}.  We find that there is some disagreement between previously published results for F$_{2}$/cc-pVQZ and our results which is illustrated in Fig.~\ref{fig:f2}.  The CIPSI results~\cite{garniron2017} were not fully converged and were not generated with orbital rotations, The previous SHCI results have large stochastic error  bars that prevent any high accuracy comparisons~\cite{sharma2017}.    Thus we are left with the FCIQMC energy of $-199.3598(2)$ Ha~\cite{cleland2012} as the best previous benchmark.
The FCIQMC result uses the initiator approximation and consequently it will be subject to systematic bias~\cite{cleland2012}.  The best ASCI result presented here has an energy of $-199.36082$ Ha using natural orbitals which is 1 mHa below the FCIQMC results.  However, we are able to validate our results with a linear extrapolation from energies generated with the Hartree-Fock orbitals.  We find the extrapolated energy agrees with our best natural orbital result to better than  0.15 mHa (see Fig.\ref{fig:f2}).  This close agreement provides numerical validation for the accuracy of the ASCI results over the more approximate results generated from CIPSI, SHCI, and FCIQMC, all of which appear to differ from ASCI by at least 1 mHa.   

We also note that simulating F$_{2}$ in the cc-pVQZ basis gave rise to some unique challenges related to orbital rotations.  In particular, compared to the other simulations presented here, it was important to use large wave functions in initial ASCI iterations, before performing the first orbital rotation for F$_{2}$. 

Table~\ref{tab:others} presents orbital rotated results for C$_{2}$ and N$_{2}$. For these systems there are no comparable results for the stochastic selected CI approaches but for C$_{2}$, there are converged DMRG results.  We see that the results from the ASCI+PT2 approach agree with the DMRG results to better than 0.25 mHa across all basis sets. For N$_{2}$, there are FCIQMC results (within the initiator approximation)~\cite{cleland2012} that can be compared, although the the initiator approximation renders comparison with ASCI less informative.

\begin{table*}

\centering
\begin{tabular}{|c|c|c|c|c|c|c|c|c|c|}
\hline
\mc{3}{}& \mc{2}{Time PT2 (secs)}  & \mc{3}{Energy (Ha)}  &  \mc{2}{ASCI speedup factor}  \\\hline
Comparisons&Dets&Basis& ASCI &SHCI & Variational & SHCI & ASCI+PT2&E(0.1mHa)& E diff(0.1mHa)  \\\hline
\rowcolor{CYAN}
C$_{2}$(8,26) HCI~\cite{sharma2017} & 28566 & cc-pVDZ &&640 &-75.7217 & -75.7286(2) &&&\\
C$_{2}$(8,26) ASCI & 10000 & cc-pVDZ &  2 && -75.71688& & -75.72805 &&\\
\rowcolor{CYAN}
C$_{2}$(8,26) ASCI & 20000 & cc-pVDZ &  3 && -75.72122& &  -75.72827 & 213& 426\\
C$_{2}$(8,26) ASCI & 100000 & cc-pVDZ &  26& & -75.72585& & -75.72852 & &  \\ 
C$_{2}$(8,26) ASCI & 182145 & cc-pVDZ &  43& &  -75.72634 & &-75.72853&& \\
\hline
\rowcolor{GREEN}
N$_{2}$(10,26) HCI~\cite{sharma2017} &37593 & cc-pVDZ && 160 & -109.2692  &  -109.2769(1) &&&\\
N$_{2}$(10,26) ASCI & 10000  & cc-pVDZ & 2 &&-109.26419 && -109.27687  && \\ 
\rowcolor{GREEN}
N$_{2}$(10,26) ASCI & 30000 & cc-pVDZ & 7 && -109.26936  & &-109.27691  & 23 & 45   \\
N$_{2}$(10,26) ASCI & 100000 & cc-pVDZ & 27 && -109.27335  & & -109.27698 &  &  \\
\hline
\rowcolor{YELLOW}
F$_{2}$(14,26) HCI~\cite{sharma2017} & 68994 & cc-pVDZ &&  11760 & -199.0913   &  -199.1001(7)  &&&\\
F$_{2}$(14,26) ASCI & 10000 & cc-pVDZ & 4 && -199.08368 && -199.09921 &&\\
\rowcolor{YELLOW}
F$_{2}$(14,26) ASCI & 100000 & cc-pVDZ &  37 && -199.09265 && -199.09929 & 310 & 620   \\
F$_{2}$(14,26) ASCI & 300000 & cc-pVDZ &  109 && -199.09406 && -199.09933 && \\
\hline
\end{tabular}
\caption{Comparison of ASCI+PT2 to HCI+PT2 for ground state energies in the cc-pVDZ basis. HCI results are taken from ref.~\cite{sharma2017}. To reach these system sizes, HCI uses a stochastic algorithm to perform the PT2 (the resulting method being called SHCI). The colors indicate which ASCI results are most comparable to the stochastic SHCI results, as determined by matching the variational energies as closely as possible.  For those comparisons we calculate the speedup that can be found by using the ASCI deterministic PT2 approach.  These ASCI speedup factors demonstrate the speedup of ASCI over stochastic SHCI both   i) for obtaining absolute energies with an expected error of 0.1 mHa with a 95\% probability "E(0.1mHa)", and ii) for obtaining the equivalent accuracy to determine a physical energy difference such as required for calculating energy gaps and atomization energies "E diff(0.1mHa)".  In simulations where stochastic and deterministic PT2 algorithms use the same variational wave function, the deterministic results will be always be more accurate because they do not have any stochastic error. The SHCI results were performed on nodes with 2 Intel Xeon E5-2680 v2 processors of 2.80 GHz,  
with 20 computational cores per node.  The timings reported for SHCI 'stochastic' are taken from  ref ~\cite{sharma2017}, for which we then calculate the single core cost (of a 2.80 GHz processor) it would take to produce 0.1 mHa accuracy with a 95\% confidence. The energies for SHCI are directly taken from reference~\cite{sharma2017}, and the reported error bars are 1$\sigma$ error bars.   The ASCI simulations were performed on a single core of a Intel Xeon E5-2620 v5 processor of 2.10 GHz. For all simulations we calculate the equivalent single core time, 
and we scale the ASCI timings to be representative of a single 2.8 GHz core. For the PT2 simulations, we 
neglect contributions less than 10$^{-8}$, as in Ref.~\cite{sharma2017}. The colors are used to highlight results that are most comparable to each other in terms of energy of the variational wave function.}
\label{tab:dz}
\end{table*}%

\begin{table*}

\centering
\begin{tabular}{|c|c|c|c|c|c|c|c|c|c|}
\hline
\mc{3}{}& \mc{2}{Time PT2 (secs)}  & \mc{3}{Energy (Ha)}  &  \mc{2}{ASCI speedup factor}  \\\hline
Comparisons&Dets&Basis& ASCI &SHCI & Variational &SHCI & ASCI+PT2&E(0.1mHa)& E diff(0.1mHa)  \\\hline
\rowcolor{CYAN}
C$_{2}$(8,58) HCI~\cite{sharma2017} & 142467 & cc-pVTZ & & 2880 & -75.7738 & -75.7846(3)&&&\\
C$_{2}$(8,58) ASCI & 50000 & cc-pVTZ &  60 && -75.768939 & &-75.784113&& \\
\rowcolor{CYAN}
C$_{2}$(8,58) ASCI & 100000 & cc-pVTZ &  117 && -75.77395 &&  -75.78447& 24& 48 \\
C$_{2}$(8,58) ASCI & 142467 & cc-pVTZ &  166 && -75.775386 & &-75.784589&& \\
\hline
\rowcolor{GREEN}
N$_{2}$(10,58) HCI~\cite{sharma2017} & 189080 & cc-pVTZ && 11520 & -109.3608  &-109.3748(6) &&&\\
N$_{2}$(10,58) ASCI & 10000  & cc-pVTZ & 19 && -109.34058 & &-109.37414 && \\ 
\rowcolor{GREEN}
N$_{2}$(10,58) ASCI & 100000 & cc-pVTZ & 184 && -109.35942  && -109.37465 &62& 124  \\
N$_{2}$(10,58) ASCI & 300000 & cc-pVTZ &  501 && -109.36519 && -109.37492&&  \\
\hline
\rowcolor{YELLOW}
F$_{2}$(14,58) HCI~\cite{sharma2017} & 395744 & cc-pVTZ &&  38880 & -199.2782 & -199.2984(9)&&&\\
F$_{2}$(14,58) ASCI & 20000 & cc-pVTZ & 60 && -199.254301 && -199.295491 &&\\
F$_{2}$(14,58) ASCI & 100000 & cc-pVTZ &  295 &&  -199.271331 && -199.296290&& \\
\rowcolor{YELLOW}
F$_{2}$(14,58) ASCI & 300000 & cc-pVTZ &  891 && -199.278140 && -199.296686 &43 & 86 \\
F$_{2}$(14,58) ASCI & 395744 & cc-pVTZ & 1163 && -199.279209 & &-199.296767 &&\\\hline
\hline
\end{tabular}
\caption{Comparison of ASCI+PT2 to SHCI for ground state energies in the cc-pVTZ basis.  SHCI results are taken from ref.~\cite{sharma2017}. See caption of Table~\ref{tab:dz} for more details. The colors are used to highlight results that are most comparable to each other in terms of energy of the variational wave function. These ASCI speedup factors demonstrate the speedup of ASCI over stochastic SHCI both   i) for obtaining absolute energies with an expected error of 0.1 mHa with a 95\% probability "E(0.1mHa)", and ii) for obtaining the equivalent accuracy to determine a physical energy difference "E diff(0.1mHa)".}
\label{tab:tz}
\end{table*}%

\begin{table*}

\centering
\begin{tabular}{|c|c|c|c|c|c|c|c|c|c|}
\hline
\mc{3}{}& \mc{2}{Time PT2 (secs)}  & \mc{3}{Energy (Ha)}  &  \mc{2}{ASCI speedup factor}  \\\hline
Comparisons&Dets&Basis& ASCI &stochastic & Variational &SHCI & ASCI+PT2&E(0.1mHa)& E diff(0.1mHa)  \\\hline
\rowcolor{CYAN}
C$_{2}$(8,108) HCI~\cite{sharma2017} & 403071 & cc-pVQZ & & 12800  & -75.7894  & -75.8018(4) &&&\\
C$_{2}$(8,108) ASCI & 10000 & cc-pVQZ &  80  && -75.75108 & &-75.799103&&  \\
C$_{2}$(8,108) ASCI & 100000 & cc-pVQZ &  670 && -75.78297 && -75.80128&& \\
\rowcolor{CYAN}
C$_{2}$(8,108) ASCI & 300000 & cc-pVQZ &  2020 && -75.79030 & &-75.80192& 6 & 12 \\
\hline
\rowcolor{GREEN}
N$_{2}$(10,108) HCI~\cite{sharma2017} & 499644 & cc-pVQZ && 64800 & -109.3884  & -109.4055(9) &&&\\
N$_{2}$(10,108) ASCI & 10000  & cc-pVQZ & 115 && -109.34421 & &-109.40349  &&\\ 
N$_{2}$(10,108) ASCI & 100000 & cc-pVQZ & 990 && -109.38073  && -109.40448 &&  \\
\rowcolor{GREEN}
N$_{2}$(10,108) ASCI & 300000 & cc-pVQZ &  2865 && -109.38782 && -109.40491 & 22&  44  \\
\hline
F$_{2}$(14,108) HCI~\cite{sharma2017} & 1053491 & cc-pVQZ && 142500 & -199.3463& -199.3590(9)  &&&\\
\rowcolor{YELLOW}
F$_{2}$(14,108) CIPSI~\cite{garniron2017} & 4000000 & cc-pVQZ &&1533503  &  -199.3417 & -199.3594   &&&\\
F$_{2}$(14,108) ASCI & 10000 & cc-pVQZ & 170 && -199.23596 && -199.35453 &&\\
F$_{2}$(14,108) ASCI & 100000 & cc-pVQZ &  1650 && -199.31893 && -199.35766 && \\
F$_{2}$(14,108) ASCI & 300000 & cc-pVQZ &  4840 && -199.32879 && -199.35843 &&\\
F$_{2}$(14,108) ASCI & 1000000 & cc-pVQZ & 15100  && -199.33765 &&-199.35913  &&\\
\rowcolor{YELLOW}
F$_{2}$(14,108) ASCI & 2000000 & cc-pVQZ & 28000  && -199.34187 &&-199.35947  &&\\
\hline
\end{tabular}
\caption{Comparison of ASCI+PT2 to SHCI for ground state energies in the cc-pVQZ basis. SHCI results are taken from ref.~\cite{sharma2017}. See caption of Table~\ref{tab:dz} for more details.  In this table we also provide a comparison to CIPSI with the stochastic hybrid approach for the F$_2$ molecule~\cite{garniron2017}.  For the CIPSI simulations, the calculation is done with a hybrid stochastic method but in the limit that all terms are calculated, and thus there is no error bar.   They were performed on a Intel Xeon E5-2680 at 2.70 GHz.  We have scaled the CPU time to 2.80 GHz which is the speed of the Intel processors for which the HCI results were calculated.  The CIPSI result requires a substantially larger number of determinants to have similar variational energies as ASCI.  We note that for cc-pVQZ F$_{2}$, the HCI wave function has a lower variational energy than ASCI for less determinants.  This is highly unusual and is the only simulation we have ever observed this property.   As a result we make direct comparisons with CIPSI, and compare the compactness of the wave function. It can be seen that the ASCI wave function with 2 million determinants has a variational energy that is comparable to the CIPSI wave function with 4 million determinants.  The colors are used to highlight results that are most comparable to each other in terms of energy of the variational wave function.}
\label{tab:qz}
\end{table*}%

\begin{table*}

\centering
\begin{tabular}{|c|c|c|c|c|c|c|c|c|c|}
\hline
\mc{3}{}& \mc{2}{Time PT2 (secs)}  & \mc{3}{Energy (Ha)}  &  \mc{2}{ASCI speedup factor}  \\\hline
Comparisons&Dets&Basis& ASCI &SHCI & Variational & SHCI & ASCI+PT2&E(0.1mHa)& E diff(0.1mHa)  \\\hline
\rowcolor{CYAN}
F$_{2}$(14,26) HCI~\cite{sharma2017} &16824  & cc-pVDZ && 3840 & -199.0871&  -199.0994(4)  &&&\\
F$_{2}$(14,26) ASCI & 10000 & cc-pVDZ & 4 && -199.08669 && -199.09927 &&\\
\rowcolor{CYAN}
F$_{2}$(14,26) ASCI & 11000 & cc-pVDZ & 5 && -199.08712 && -199.09928 & 768& 1536\\
F$_{2}$(14,26) ASCI & 100000 & cc-pVDZ & 42 && -199.09510 &&  -199.09930 & &\\
\hline
\rowcolor{GREEN}
F$_{2}$(14,58) HCI~\cite{sharma2017} & 141433 & cc-pVTZ &&23520 & -199.2787 &  -199.2972(7)  &&&\\
F$_{2}$(14,58) ASCI & 10000 & cc-pVTZ & 42 && -199.25358 &&  -199.29831 &&\\
\rowcolor{GREEN}
F$_{2}$(14,58) ASCI & 100000 & cc-pVTZ &  390 && -199.27901 && -199.29660 &60&120\\
F$_{2}$(14,58) ASCI & 150000 & cc-pVTZ &  590 && -199.28110 &&  -199.29667 &&\\
F$_{2}$(14,58) ASCI & 300000 & cc-pVTZ &  1036 && -199.28523&& -199.29648 &&\\
\hline
\rowcolor{YELLOW}
F$_{2}$(14,108) HCI~\cite{sharma2017} & 221160 & cc-pVQZ &&174960  & -199.3355 &  -199.3590(9) &&&\\
\rowcolor{YELLOW}
F$_{2}$(14,108) ASCI & 150000 & cc-pVQZ &  2870 && -199.33462  &&-199.36076 &60&120\\
F$_{2}$(14,108) ASCI & 200000 & cc-pVQZ &  3690 && -199.33724  &&-199.36080 &&\\
F$_{2}$(14,108) ASCI & 300000 & cc-pVQZ &  5340 && -199.34072  &&-199.36083 &&\\
F$_{2}$(14,108) ASCI & 1000000 & cc-pVQZ &  15200 && -199.34609  &&-199.36084 &&\\
F$_{2}$(14,108) ASCI & 2000000 & cc-pVQZ &  28800 && -199.34992  && -199.36082 &&\\
\hline
\end{tabular}
\caption{Comparison of ASCI+PT2 to stochastic SHCI for ground state energies of F$_{2}$ with orbital rotations turned on. HCI results are taken from ref.~\cite{sharma2017}. The deterministic ASCI+PT2 approach has significantly improved performance over stochastic SHCI, compared to the results without orbital rotations in Tables~\ref{tab:dz}, ~\ref{tab:tz} and ~\ref{tab:qz}.  The timings reported for SHCI 'stochastic' are taken from  reference~\cite{sharma2017}, for which we then calculate the single core cost (of a 2.80 GHz processor) it would take to produce 0.1 mHa accuracy with a 95\% confidence.  
 The energies for SHCI are directly taken from reference~\cite{sharma2017}, and the reported error bars are 1$\sigma$ error bars.  The colors are used to highlight results that are most comparable to each other in terms of energy of the variational wave function.}
\label{tab:orbrot}
\end{table*}%

\subsection{TZV basis calculation for Cr$_{2}$}

The Cr$_{2}$ dimer in the SVP basis set (at 1.5\AA) has provided a standard benchmark for
testing the efficiency of strongly correlated methods~\cite{yanai2009,amaya2015, tubman2016-1,holmes2016,wirawan2015}.  To date, only DMRG and selected CI techniques, such as ASCI,
have been able to converge the result to below 1 mHa accuracy.  Benchmarks have recently been generated with semi-core simulations using at least 24 electrons with the Dunning basis sets (CIPSI)~\cite{garniron2017}, and with DMRG using DKH corrections with a triple zeta basis~\cite{guo2018-1,guo2018-2}.  
 Here we present results for Ahlrich triple zeta valence (TZV) basis set~\cite{svp} (24e, 76o), which 
is expected to be manageable to converge for most selected CI implementations and for DMRG.   Computation of our benchmark energies presented here takes only a few hours, as shown in Table~\ref{tab:cr2bench}.  The ASCI simulations were performed on a single core of a Intel Xeon E5-2620 v5 processor of 2.10 GHz.

\begin{table*}
\centering
\begin{tabular}{|c|c|c|c|c|c|c|c|c|}
\hline
 \mc{3}{ }  &\mc{2}{Contributions (billions)} & & \mc{3}{Energy (Ha)}  \\\hline
\hline
Comparisons&Dets&Basis&Total &cutoff 10$^{-8}$ & Unique $\alpha$&  Variational&ASCI+PT2 & Reference \\\hline
C$_{2}$(8,26) & 10000 & cc-pVDZ &0.05 & 0.006 &1096 & -75.72289 & -75.72857 &\\
C$_{2}$(8,26)& 100000 & cc-pVDZ &0.5 & 0.06 &4381 &-75.72785 & -75.72855 & \\ 
C$_{2}$(8,26) & 300000 & cc-pVDZ &1.6 & 0.18 &7401 &-75.72836 & -75.72855 & \\ 
DMRG~\cite{amaya2015} &  &  & & & & &&-75.72855  \\ 
\hline
C$_{2}$(8,58) & 10000 & cc-pVTZ &0.3 & 0.08 &1378 &-75.76592 &-75.78513 & \\
C$_{2}$(8,58) & 100000 & cc-pVTZ &3.2 & 0.7 & 6992 &-75.77917 & -75.78518 & \\
C$_{2}$(8,58) & 300000 & cc-pVTZ &9.6 &1.8 &14380 & -75.78196 & -75.78515 &\\
DMRG~\cite{amaya2015} &  &  & & & & &&-75.785054  \\ 
\hline
C$_{2}$(8,108) & 10000 & cc-pVQZ & 1.2 & 0.5 & 1436&-75.77663 &-75.80409& \\
C$_{2}$(8,108) & 100000 & cc-pVQZ &11.9 &4.7 & 8194 & -75.79335 & -75.80313& \\
C$_{2}$(8,108) & 300000 & cc-pVQZ &35.7 & 9.7 & 18692 & -75.79807 & -75.80290&  \\
DMRG~\cite{amaya2015} &  &  & & & & &&-75.802671  \\ 
\hline
N$_{2}$(10,26) & 10000 & cc-pVDZ &0.07&0.01 & 1201 &-109.26837& -109.27708&  \\
N$_{2}$(10,26) & 100000 & cc-pVDZ &0.7 &0.1 &5447 &-109.27522 &-109.27699 & \\
N$_{2}$(10,26) & 300000 & cc-pVDZ &2.3 & 0.2 & 11069&-109.27638 & -109.27699&  \\
FCIQMC~\cite{cleland2012} &  &  & & & & &&-109.2767(1) \\ 
\hline
N$_{2}$(10,58) & 10000 & cc-pVTZ &0.5 &0.1 & 1495 & -109.35150 & -109.37640 & \\
N$_{2}$(10,58) & 100000 & cc-pVTZ &4.9 & 1.1 & 7751 &-109.36705&-109.37661 &  \\
N$_{2}$(10,58) & 300000 & cc-pVTZ & 14.7& 3.0 & 17060 &-109.37041 & -109.37655 & \\
N$_{2}$(10,58) & 600000 & cc-pVTZ & 29.5& 5.1 & 25955 &-109.37192 & -109.37643 & \\
FCIQMC~\cite{cleland2012} &  &  & & & & &&-109.3754(1)  \\ 
\hline
N$_{2}$(10,108) & 10000 & cc-pVQZ &1.8 & 1.5& 1568 &-109.35964 & -109.40842 & \\
N$_{2}$(10,108) & 100000 & cc-pVQZ &18.6 & 13.6 &8399 &-109.38871 & -109.40639& \\
N$_{2}$(10,108) & 300000 & cc-pVQZ &55.7 &28.4 &19435 &-109.39517 & -109.40617 &  \\
FCIQMC~\cite{cleland2012} &  &  & & & & &&-109.4058(1)  \\ 
\hline
\end{tabular}
\caption{ASCI+PT2 with orbital rotations for C$_{2}$ and N$_{2}$.  For C$_2$, comparisons are made to DMRG~\cite{amaya2015}. For N$_{2}$, comparisons are made to FCIQMC~\cite{cleland2012}.  The DMRG results represent a highly converged benchmark, whereas the FCIQMC results are approximate, with uncontrolled errors deriving from the initiator approximation.  As a result, the best ASCI results agree with DMRG to better than 0.25 mHa, while agreement with the approximate FCIQMC results is not quite as good. These calculations show much quicker convergence (with respect to number of determinants) than the respective simulations without orbital rotations, as expected.  The column ``Contributions" lists how the number of PT2 contributions, as well as the size of the subset that have values greater than 10$^{-8}$.  The column ``Unique $\alpha$" lists the number of unique $\alpha$ bitstrings in the variational wave function.}
\label{tab:others}
\end{table*}

\begin{table*}
\centering
\begin{tabular}{|c|c|c|c|c|c|c|c|}
\hline
Cr$_{2}$(24e,76o)& \mc{2}{ PT2 times (secs)}  &\mc{3}{Contributions (billions)} & \mc{2}{Energy (Ha)}  \\\hline
Dets& Total & Sort &Total & > 10$^{-8}$ & Unique $\alpha$  &Variational  &ASCI+PT2 \\\hline
10000& 112 & 25 & 2.8 & 0.38 & 1649 &  -2086.71750 & -2086.95370\\
100000& 1456 &512 &28 & 6.5 & 9897 &-2086.821072 &-2086.93277 \\
300000& 4050 & 1300 & 84 &15 &26548 &-2086.84799 & -2086.93355  \\
1000000&10625 & 2755 & 282 & 32 &68488 & -2086.87156 &-2086.93361   \\
\hline
\end{tabular}
\caption{Benchmark ASCI+PT2 calculations for the Cr$_{2}$ molecule in the Ahlrich triple zeta basis set. Similar to the SVP benchmarks~\cite{tubman2016-1,amaya2015}, we use 24 active electrons.  In total there are 76 spatial orbitals.  The column ``PT2 times'' lists the total time for the PT2 calculation (Total), as well as the amount of time spent in sorting (Sort).   The column ``Contributions" lists how the number of PT2 contributions (including duplicates), and the size of the subset of contributions greater than 10$^{-8}$.  The column ``Unique $\alpha$" lists the number of unique $\alpha$ bitstring in the variational wave function.}
\label{tab:cr2bench}
\end{table*}

\section{Speed improvements to the naive deterministic algorithm}

The naive deterministic approach presented in Algorithm \ref{alg:constraint} will be quite fast even in the absence of substantial optimizations. A number of further speedups to this naive approach are however possible. In this section we describe some of the techniques that we have developed to accelerate the constraint PT2 approach and present the  substantially faster run times achieved from these additional refinements.   

The techniques described here are:
\begin{itemize}
\item Fast generation of triplet contributions
\item Removing duplicates with the core determinants
\item Spin symmetry (Z$_{2}$) for $S_{z}=0$ calculations
\item Data reuse for unique $\alpha$ and $\beta$ strings (triplet constraint specific)
\item Matrix element cutoff and fast diagonal matrix elements
\item Parallelization with work lists.
\item Generalized compression with hash functions
\end{itemize}
A GPU could also be employed to reduce the sorting time even further.  We generally find that a single GPU is about 10 times faster than a single CPU for sorting
~\cite{tubman2018}.  There is still much ongoing development of GPU sorting~\cite{gpusort2017}.  Much of the current runtime is used to move data onto and off the GPU, so a fully GPU implementation could have even further speed improvements.  

\subsubsection{Fast generation of triplet contributions}
  One of the main costs for the PT2 algorithm presented in this work is the cost for generating the contributions.  For every constraint, one has to loop over all determinants.  Thus the constraint generation routine will be called $N_{tdets}$ times for each triplet constraint considered.   The most expensive  part of generating the constrained  PT2 contributions is creating the contributions from the double alpha or double beta excitations.  To create these excitations quickly, we consider three cases for creating a list of pair of occupied orbitals and three cases for creating a list of pairs of virtual orbitals.  Once these lists are created, every pairwise combination generates all excitations from D$_{i}$ that satisfy the triplet constraint.  The different cases are described in Algorithm~\ref{alg:triplet}.

\begin{figure}[tpb]
\begin{algorithm}[H]
 \begin{algorithmic}[1]
\State For each determinant $D_{i}$ in the trial wave function, generate only the relevant PT2 contributions that match the current triplet string $T$. We show the more complicated case of double excitations here.  Do this by creating a reduced list of occupied pairs ($\{O\}$) and virtual pairs  ($\{V\}$)  as follows:
\State Notation: $\land$ is (bit-wise) logical AND, $\oplus$ is (bit-wise) logical XOR
\State Definition $T_{small}$: The smallest occupied in the triplet $T$  
\State Create a bitmask $B$, which is equal to 1 for all orbitals greater than $T_{small}$.
\State  If countbits$(D_{i}\land T)$ = 0 or countbits$((D_{i} \land B) \oplus T)$ > 2, return empty ($\{O\}$)  and ($\{V\}$) (There are too many differences between $D_i$ and $T$ to be fixed with a double excitation)
\State  Begin case for creating $\{V\}$:
\begin{itemize}
\item Case(1):  If countbits$(D_{i} \land T)$ == 1, put the pair of orbitals that correspond to the bits in $((D_{i} \oplus T) \land T)$ into  ($\{V\}$) 
\item Case(2):  If countbits$(D_{i} \land T)$ == 2, put the pairs $(x,((D_{i} \oplus T) \land T))$ into  ($\{V\}$), where $x$ is any unoccupied orbital smaller than $T_{small}$
\item Case(3):  If countbits$(D_{i} \land T)$ == 3, put the pairs $(x,y)$ into  ($\{V\}$), where $x$ and $y$ are distinct unoccupied orbitals smaller than $T_{small}$
\end{itemize}
\State  Begin case for creating $\{O\}$:
\begin{itemize}
\item Case(1):  If countbits$((D_{i} \land B) \oplus T)$ == 2, put the pair of orbitals that correspond to the bits in $(D_{i} \land B) \oplus T)$ into  ($\{O\}$) 
\item Case(2):  If countbits$((D_{i} \land B) \oplus T)$ == 1, put the pairs $(x,((D_{i} \land B) \oplus T))$  into  ($\{O\}$), where $x$ is any occupied orbital less than $T_{small}$
\item Case(3):  If countbits$((D_{i} \land B) \oplus T)$ == 0, put the pairs $(x,y)$ into  ($\{O\}$), where $x$ and $y$ are distinct occupied orbitals smaller than $T_{small}$
\end{itemize}
\State Double loop over the elements of ($\{O\}$)  and ($\{V\}$)  to create all possible double excitations from $D_{i}$ with the orbitals in $T$ as the highest occupied.
 \end{algorithmic} 
 \caption{Generate fast triplet contributions}
 \label{alg:triplet}
\end{algorithm}
\end{figure} 

\subsubsection{Removal of duplicate PT2 contributions from the variational wave function}
Only determinants absent from the variational SCI wave function contribute to the PT2 energy. However the list of all possible single and double excitations 
out of the variational SCI wave function also contains all the determinants present in the SCI wave function (since the search algorithms should guarantee that any determinant in the variational wave function is connected to at least one other determinant in the variational wave function). This situation persists even after imposition of a constraint.  Therefore measures must be taken to avoid contamination of the PT2 energy from the variational wave function determinants. 

A simple approach to this would be to explicitly check whether a given excitation is present in the variational SCI wave function or not. However, this is quite inefficient, with a cost of verification scaling at least as $\log N_{tdets}$ for each excited state determinant (assuming an efficient algorithm such as binary search is used), and there would be $N_{SD}N_{tdets}$ such excitations. The net cost of this simple approach would therefore show a rather prohibitive $N_{SD}N_{tdets}\log N_{SD}$ scaling. 

A more efficient approach is to calculate the PT2-like contributions from determinants present in the variational wave function in advance, and subtract these from the PT2-like energy computed from all possible excitations out of the variational wave function. This would yield a PT2 energy without any contamination from the determinants of the variational wave function. The PT2-like contribution from determinants in the variational wave function is calculated efficiently with only a linear $N_{tdets}$ scaling as follows:
\begin{enumerate}
\item For true PT2 contributions $\ket{D_{i}}$ that are not included in the variational wave function $\ket{\psi}$ (i.e., contributions not in the variational space), we attempt to find $\bra{D_i}H\ket{\psi}$ by computing contributions from all variational wave function determinants that are a single or double excitation away from $\ket{D_{i}}$. This is exact if $\ket{D_{i}}$ itself is not in the variational wave function.

\item If $\ket{D_{i}}$ was in the variational wave function, with coefficient $c_i$, we instead calculate $\bra{D_i}H\ket{\psi}-c_i\bra{D_i}H\ket{D_i}$, where the negative term arises because the algorithm finds single and double excitations only, which misses the zero excitation diagonal term $c_i\bra{D_i}H\ket{D_i}$.

\item  To proceed, recall that $\bra{D_i}H\ket{\psi}=E_{ASCI}\braket{D_i}{\psi}=c_iE_{ASCI}$, since $\ket{D_{i}}$ and  $\ket{\psi}$ both belong to the subset of the Hilbert space spanned by determinants employed in the variational wave function, and $\ket{\psi}$ is furthermore an eigenstate of $H$ within that subspace. 
\item 
Construct a contribution of $\bra{D_i}H\ket{\psi}-c_i\bra{D_i}H\ket{D_i}=c_i(E_{ASCI}-H_{ii})$ for each $\ket{D_{i}}$ in the variational wave function. This corresponds to a PT2 like contribution of $\dfrac{|\bra{D_i}H\ket{\psi}-c_i\bra{D_i}H\ket{D_i}|^2}{E_{ASCI}-H_{ii}}=c_i^2(E_{ASCI}-H_{ii})$. 
\item These extra PT2 like contributions $c_i^2(E_{ASCI}-H_{ii})$ are summed over all $N_{tdets} $ $\ket{D_{i}}$'s, and are 
then subtracted from the PT2-like energy computed by the algorithm, yielding the true PT2 energy.
\end{enumerate}

\subsubsection{Spin symmetry (Z$_{2}$) for $S_{z}=0$ calculations}
  For systems in which the quantum number for S$_{z}$ = 0,
one can loop over all determinants at the end of an ASCI run and ensure that, for all $\alpha$/$\beta$ pairs, the bitstring corresponding to the swap of the $\alpha$/$\beta$ strings is also present.  If not, the corresponding bitstring can be added in and the wave function re-diagonalized (large ASCI wave functions generally are very close to having this symmetry automatically). When the ASCI wave function has this $\alpha$/$\beta$ symmetry, then the set of determinants contributing to the PT2 correction will also have this symmetry, and roughly half of the contributions can be ignored.  This can be implemented in practice by only generating PT2 contributions for which the $\alpha$ bitstrings are not smaller than the $\beta$ bitstrings.  
These contributions are then multiplied by 2 to account for the excluded cases.

\subsubsection{ Data reuse for unique $\alpha$ and $\beta$ strings (triplet constraint specific)}

We can also sort the ASCI wave function by the numerical $\alpha$ bitstrings prior to calculating the PT2 contribution.  This permits reuse of information and calculations associated with a given $\alpha$ bitstring.  For example, the double $\alpha$ excitations will be the same for all bitstrings with the same $\alpha$ bitstring, and the off-diagonal Hamiltonian matrix elements will also be the same for these excitations.  Thus it is possible to reuse the generation of PT2 contributions and matrix element calculations over all reference determinants that differ only in the beta bitstring.  This is a significant
speedup, since the number of unique $\alpha$ bitstrings is often substantially smaller than the size of the variational wave function, and by percentage, goes down as the latter gets bigger.  For virtually all simulations calculated here and in previous ASCI work~\cite{tubman2016-1,tubman2018}, the number of unique $\alpha$ bitstrings is generally less than 10\% of the total number of variational determinants for wave functions of size 100,000.  This often becomes less than 5\% when using up to 1 million determinants.  In Table~\ref{tab:others} and ~\ref{tab:cr2bench}, we demonstrate this trend by presenting the number of unique $\alpha$ bitstrings for each calculation.

\subsubsection{Matrix element cutoff and fast diagonal matrix elements} 

   During the above process, when calculating matrix elements, a cutoff is used for determining whether a given matrix element should be stored and used. For the simulations considered here, we generally use a cutoff of 10$^{-8}$ for inclusion of matrix elements in the PT2 calculations.  As noted above, a cutoff is not required, but it can slightly speed up the algorithm without any significant loss of accuracy for the calculations under consideration here.  We then calculate the diagonal matrix element in the denominator of Eq.~\ref{eqn:en} only after checking that the matrix element in the numerator is larger than the cutoff value.  For evaluation of the diagonal matrix elements, we use the fast diagonal matrix element algorithm that was presented in ref.~\cite{tubman2018} and is also included here in the appendix.

\subsubsection{ Parallelization with work lists }

The constraint PT2 algorithm presented in this work can be calculated in parallel over the constraints without any communication between parallel units, as described in previous sections.  After the contributions have been generated, it is possible to do a parallel sort over the contributions or, alternatively, to offload the sorting to a GPU.  We have previously presented benchmarks for the parallelization and sorting with the Thrust library~\cite{thrust} and the IPS$^{4}$O parallel sort~\cite{ips4}.  The amount of work for each triplet constraint scales with the number of PT2 contributions that are consistent with the constraint.  Before calculating any of the PT2 contributions, it is possible to do a first pass over the ASCI wave function and determine how many contributions will  be consistent with each triplet.  This work list can be saved and used to make sure that enough memory is available to calculate the contributions for each triplet constraint. It can also be used to provide an efficient load distribution for parallel execution.

\subsubsection{Generalized compression with hash functions}
\label{sec:hash}
Going to larger basis sets becomes more costly for the ASCI approach, for several reasons. One of the biggest cost increases is due to the manipulation of larger bitstrings associated with a larger basis set.   Some of this cost can be mitigated by using more compact bitstring representations. While the standard bitstring representation is suitable for non-selected CI algorithms, there might be better forms for SCI approaches. 

Here we consider how hashing can be used to compress bitstrings for use in calculating the PT2 energy.  Hashing is an extremely fast way to compress a bitstring. The one negative feature of hashing, namely that the compression can induce collisions, i.e., two long bitstrings can be compressed to the same shorter value, can be managed by taking advantage of hash functions designed to avoid  certain types of collisions.  For example many  hash functions have the property called avalanching, where small changes of any bit in the uncompressed bitstring can lead to a larger difference in the hashed string.   Such a feature is important for our purposes and is also well tested and understood for all widely used hash functions~\cite{smhasher1,smhasher2}. Compression of bitstrings for selected CI applications is  highly desirable and will very generally provide speedups when used in a sorting/hashing approach.

It is important to understand how often collisions may occur when hashing.  As an example, 
there are over 10$^{36}$ unique numbers that can be represented with a 128 bit integer, which is the length of bitstrings used in the SVP Cr$_{2}$ calculation. Yet, at most, only on the order of 10$^{13}$ PT2 contributions will be generated in a very large scale PT2 calculation.  If we consider compressing 128 bitstrings to bitstrings of length 64, we expect that a collision is highly unlikely for applications of selected CI simulations. That is to say, the number of determinants being considered will not be dense in comparison to the number of values that can be represented with 64 bits (which is $\approx 10^{18}$). 

\begin{table*}
\centering
\begin{tabular}{c|c|c|c|c|c|c|c|}
\hline
 & FNV1a-YT (32) & t1ha(64) & Spooky(32) & Spooky(64) & Murmur3a(32) & xxhash(32) & xxhash(64) \\
\hline
128 bit integers & 20.00 & 32.00 & 68.00 & 68.00 & 41.00 & 36.00 & 48.00 \\
256 bit integers & 32.00 & 35.00 & 77.00 & 76.00 & 54.00 & 42.00 & 70.00 \\
\end{tabular}
\caption{A test of cycles per hash with different hash functions (smaller is better). Tested on a linux machine intel 2.4 GHZ. Compilation and testing was done with the SMhasher testing suite~\cite{smhasher1,smhasher2}. We present here results from a subset of the hash functions included~\cite{xxhash,t1ha,murmur,spooky,fnv}. Different hash functions tested are either of the 32 bit or 64 bit variety. We performed tests on input data sizes of 128 bit and 256 bits. For ASCI simulations we have been performing most of our tests with 64 bit hash functions, and the t1ha hash function. 
Shown here are results for a mix of popular hash functions, including the best performing functions.  Performance tests with other hash functions on other machines are available~\cite{smhasher2,hashtest1}}
\label{tab:hashspeed}
\end{table*}

For testing we considered \textit{64 bit} hash functions.  The term \textit{64 bits} indicates the size of the output integer that is created when a longer bitstring is hashed. For the electronic structure applications considered here, we expect that sorting up to a billion numbers at once will be roughly the high end of what will be needed. 
As an example, consider the ASCI sorting algorithm for purposes of calculating a PT2 energy. For many of the simulations done in this work, we might expect to have to sort 100 million PT2 contributions, each of 128 bits. If we were to hash these PT2 contributions, with a 64 bit hash, the probability of a hash function producing no collisions over 10$^{8}$ elements is approximately given by the formula $1-e^{-k^{2}/2N}$, where $k$ is the number of hashed values, and $N$ is the number of elements in a 64 bit integer.  For a 64 bit integer, $N\approx 10^{18}$. Thus, in this example, there is a greater than 99\% chance that there are no collisions. Hashing to 32 bit integers on the other hand, would likely result in many collisions although it is possible to test for collisions.  Such an approach might be considered if it eventually became much faster to sort 32 bit integers relative to 64 bit ones, which is currently not the case with our tests.   

The speed at which hashing occurs can vary among hash functions. In Table \ref{tab:hashspeed} we present test data of the average speed per hash for both 128 bitstrings and 256 bitstrings over a few well known 32 and 64 bit non-cryptographic hash functions.  The tests were performed with hash implementations from the following repository~\cite{smhasher1,smhasher2}.   We find that hashing provides significant speedups when going to large basis set calculations.  For example we see a factor of two speedup with cc-pVQZ simulations of C$_{2}$ (and other cc-pVQZ simulations) when hashing bitstrings to 64 bits before sorting.   Different sorting algorithms have been discussed and tested in our previous work~\cite{tubman2018}.

\section{Conclusions}
In this work, we have presented an efficient deterministic alternative for the second order perturbation theory (PT2) refinement to a selected CI method. This algorithm leverages a combination of fast sorting algorithms on modern computers with a well thought out algorithmic design. The result is a highly efficient algorithm that is in many cases two orders of magnitude faster than the recently proposed stochastic approaches to evaluation of PT2 corrections.  This approach allowed us to converge the ground state energies of several different molecules which previous selected CI approaches have had difficulties converging, including F$_{2}$.  
It might also be possible to use the approaches developed here to produce a more efficient stochastic PT2 algorithm, though the lack of stochastic error would ensure that any deterministic algorithm would remain arbitrarily more accurate (ignoring numerical precision errors that would be present in both approaches). 

We also found that the SCI+PT2 energy converges faster when a more compact variational wave function is used as the reference. Such wave functions can be readily obtained with the ASCI algorithm, and compactness can further be enhanced through natural orbital rotations. All these improvements to efficiency of the ASCI approach and its PT2 refinement, leads to the conclusion that ASCI is well-suited to pursue many different electronic structure problems in chemistry and physics that have been difficult to pursue with other techniques and approaches. 

\section{Acknowledgements}
This work was supported through the Scientific Discovery through
Advanced Computing (SciDAC) program funded by the U.S. Department of
Energy, Office of Science, Advanced Scientific Computing Research and
Basic Energy Sciences.  We used the Extreme Science and Engineering Discovery
Environment (XSEDE), which is supported by the National Science Foundation Grant No. OCI-1053575.  DH was supported by a Berkeley Fellowship. DSL and MHG acknowledge support from the Director, Office of Science, Office of Basic Energy Sciences, of the U.S. Department of Energy under Contract No. DE-AC02-05CH11231.

\section{Appendix:}
\subsubsection{Fast Diagonal Matrix Elements}
\label{sec:fdiag}

During the calculation of the denominator of equation \ref{eqn:rankeqn}, the diagonal matrix element of the connection being considered is required. Calculating the diagonal matrix elements is a relatively costly step because these elements involve sums over both the number of electrons and the number of pairs of electrons. However, because the determinant whose diagonal is sought is always a single/double excitation away from a reference determinant being searched from, the denominator can be calculated quickly for all connections from the reference determinant. Consequently, E$_{ref}$ - E$_{sd}$ only involves a small subset of terms and can be calculated much faster than E$_{sd}$ could be evaluated \textit{ab initio}.  Algorithm~\ref{alg:diag} describes the protocol. Briefly, the contributions to the energy of the excited electrons are calculated and added to the reference determinant energy, while the contribution of the removed electron is subtracted. Calculating the array of partial contributions described in Algorithm~\ref{alg:diag} has to be done only once per reference determinant. 

\begin{figure}
\begin{algorithm}[H]
 \begin{algorithmic}[1]
\State (Precalculation step) Input Determinant $D_{i}$, the diagonal matrix element $H_{ii}$, and the one-electron integrals $h_{ii}$
\State (Precalculation step) Calculate the partial contribution: $p(i) = \sum^{occ}_{j} \langle ij||ij\rangle$
\State Input $D_k$ (connected to $D_i$) with set of orbitals excited into ($A$) and the set of orbitals excited out of ($R$)
\State $E_{rem} = \sum_{i\in R}h_{ii}+p(i)$, $E_{add} = \sum_{i\in A}h_{ii}+p(i)$
\State $H_{kk}=H_{ii}-E_{rem}+E_{add}-\sum_{i\in R, j\in A}\langle ii||jj\rangle$
\If{$A, R$ have two elements from the same spin space}
$H_{kk} = H_{kk} + \langle R_{1}R_{1}||R_{2}R_{2}\rangle-\langle A_{1}A_{1}||A_{2}A_{2}\rangle-\langle R_{1}R_{1}||A_{2}A_{2}\rangle-\langle R_{2}R_{2}||A_{1}A_{1}\rangle$
\ElsIf{$A, R$ have two elements from different spin spaces}
$H_{kk} = H_{kk} + \langle R_{\alpha}R_{\alpha}|R_{\beta}R_{\beta}\rangle+\langle A_{\alpha}A_{\alpha}|A_{\beta}A_{\beta}\rangle-\langle A_{\alpha}A_{\alpha}|R_{\beta}R_{\beta}\rangle-\langle R_{\alpha}R_{\alpha}|A_{\beta}A_{\beta}\rangle$
\EndIf
 \end{algorithmic} 
 \caption{Fast Diagonal Matrix elements}
 \label{alg:diag}
\end{algorithm}
\end{figure}


%
\end{document}